\documentclass{entcs}
\usepackage{alltt}
\usepackage{entcsmacro}
\usepackage{graphicx}
\usepackage{amsfonts,amsmath,amssymb}
\usepackage{verbatim}
\usepackage{latexsym}

\def\lastname{Bertot and Komendantskaya}
\begin{document}
\begin{frontmatter}
  \title{Inductive and Coinductive Components of Corecursive Functions in Coq} 
  \author{Yves Bertot \thanksref{ALL}\thanksref{myemail}}
  \address{Project MARELLE\\ INRIA Sophia Antipolis\\
    France} 
  \author{Ekaterina Komendantskaya\thanksref{ALL}\thanksref{coemail}}
  \address{Project MARELLE\\INRIA Sophia Antipolis\\
    France} 
 \thanks[ALL]{The work was funded by the INRIA CORDI post-doctoral program and
the ANR project ``A3Pat'' ANR-05-BLAN-0146.} 
  \thanks[myemail]{Email: \href{mailto:yves.bertot@inria.fr}{\texttt{\normalshape
      yves.bertot@inria.fr}}} 
  \thanks[coemail]{Email: \href{mailto:ekaterina.komendantskaya@inria.fr}{\texttt{\normalshape ekaterina.komendantskaya@inria.fr}}}
\begin{abstract}
In Constructive Type Theory, recursive and corecursive definitions are subject to syntactic restrictions which guarantee termination for recursive functions and productivity for corecursive functions.
However, many terminating and productive functions do not pass the syntactic tests. Bove proposed in her thesis an elegant  reformulation of the method of  accessibility predicates that 
widens the range of terminative recursive functions formalisable in Constructive Type Theory. In this paper, we pursue the same goal for productive corecursive functions. Notably, our method of formalisation of 
coinductive definitions of productive functions in Coq requires not only the use of ad-hoc predicates, but also a systematic algorithm that separates the inductive and coinductive parts of functions.
\end{abstract}
\end{frontmatter}

\section{Introduction}

The proof assistant Coq \cite{Coq} is an implementation of the Calculus of Inductive Constructions \cite{CH88} extended with inductive \cite{PM93} 
and coinductive \cite{Gim96} types.
Implementations of coinductive types were first suggested by Coquand in \cite{Coq94} and implemented in Coq  by Gimenez \cite{Gim96}. 
Coq has proved to be an effective tool for working with different kinds of final coalgebras, such as the final coalgebra of streams \cite{BC04}, 
the final coalgebra of infinite binary trees  \cite{BC04}, and some others.
For instance, the machinery of Coq was used to define algebraic structures on real numbers \cite{DG93,Niq06,Bert07}.

The specification language of Coq makes it possible to model the 
types and programs of typed programming languages.
Typed programming languages usually provide a few basic types and a mechanism that allows the definition of \emph{inductive} data types.
When defining an inductive data type, we need to introduce constructors to generate the elements of the new type.
A very well-known inductive data type is the type of natural numbers, defined using two constructors: $0$ and $S$.

In Coq, one can define coinductive types in the same fashion as inductive types, using a few basic constructors that are also related to destructors through the
pattern-matching construct. One can also use destructors 
when he wants to emphasise the duality relative to  inductive constructors.
We will illustrate this in Section \ref{sec:indcoind}.

A key notion in typed (functional) programming is the notion of \emph{recursion} by  which an object being defined refers to itself.
Functions defined over inductive types are recursive by nature. E.g., most of the functions defined on natural numbers need to be defined recursively. 
In Section \ref{sec:indcoind}, we will discuss in more detail 
the syntax of such functions in Coq.

Associated with recursion, there is a crucial question of termination. In general, there is no guarantee that a recursive function will always terminate.
A solution to this problem is to use only \emph{structurally recursive} definitions. A structurally recursive definition is such that every recursive call is 
performed on a structurally smaller argument.
This guarantees that the recursion terminates. Constructive Type Theory in general, and Coq in particular, impose the structurally recursive condition on every defined recursive function. 
And thus, all functions are guaranteed to terminate in Coq, as we further explain in Sections \ref{sec:termprod} and \ref{sec:strguard}. 

Definitions where the recursive calls are not required
 to be on structurally smaller arguments, that is, where the recursive calls can be performed on any argument are called \emph{general recursive} definitions. 
Many important and well known algorithms are not structurally recursive but general recursive. Although many general recursive algorithms can be proved to terminate, 
there is no \emph{syntactic condition} that guarantees their termination and thus, general recursive algorithms have no direct formalisation in Constructive Type Theory and in Coq.  
Several solutions to the problem of encoding general recursive functions have been suggested in \cite{Abel06phd,BFGPU04,Cap05}.
Other methods include the use of \emph{accessibility predicates} \cite{Acz77,BB00,Nord88} and \emph{ad-hoc predicates} \cite{Bov01}, the two latter methods are surveyed in Section \ref{sec:strguard}.
Most of these methods ultimately rely on structural recursion.

Already in the case of inductive types, there exists a difference between
the class of functions satisfying the  \emph{semantic condition} of
termination, and the class of functions satisfying the
\emph{syntactic condition} of structural recursion. It is significant
that, when working with coinductive types in Coq, we find the notions
of productivity \cite{Dij80,Sij89} and guardedness \cite{Coq94,Gim96}
dual to those of termination and structural recursion. This is analysed in  Sections \ref{sec:termprod} and \ref{sec:strguard}.
Similarly to the inductive case, the guardedness condition bans Coq formalisations for many useful productive functions. 
This problem was tackled (e.g.) in \cite{Abel06phd,BFGPU04,GM02,GM04} for type theory and
in \cite{Mat99} for HOL.

A particular application of an ad-hoc predicate for defining
corecursive filter function on streams in Coq first appeared in \cite{Bert05}, where
it was used to formalise Eratosthenes' sieve.
Already in that  example, the filter function was decomposed
into inductive and coinductive components, in order to become guarded. 
A similar result, also for filter functions on streams, was
described by M.~Niqui \cite{Niq04,Niq06}.

However, neither \cite{Bert05} nor \cite{Niq04,Niq06} included the
systematic description of the method of separating inductive and
coinductive components of productive values in general case. The
general description of how the ad-hoc predicates can be constructed
for corecursive functions was missing, too.

In this paper we venture to generalise the results obtained in \cite{Bert05}, 
and describe the method of formalising productive (non-guarded) values in 
Coq for any given function and for any given data type. In Section 
\ref{sec:strguard} we describe the class of functions that are covered 
by this general method; and give counterexamples of functions that are not.

In
Sections \ref{sec:coadhoc} and \ref{sec:coind}, we 
give a general method of separating the inductive and coinductive components of productive coinductive values in Coq.
In particular, Section \ref{sec:coadhoc} is devoted to the inductive component and 
gives general characterisation of the
method of building ad-hoc predicates for formalising productive values in Coq.
In Section \ref{sec:coind}, we characterise the coinductive component. 

Notably, there are two predicates, \texttt{eventually} and
\texttt{infinite}, that are essential for characterisations of
inductive and coinductive components, respectively.  The similar ``eventually'' and
``infinite'' were first introduced as temporal modalities in
\cite{Pnu81}, and their coalgebraic specification was given by Jacobs
in \cite{Jac02}.  We show how to formalise lemmas relating \texttt{eventually} and
\texttt{infinite}, and use these lemmas to tie together the
inductive and coinductive components of productive functions.

In Section \ref{sec:eqlemma}, we prove the ``recursive equation lemmas'' establishing 
that our formalisations of the productive functions are correct. Prior to this paper, such lemmas have never been established; 
in particular, they were missing in \cite{Bert05,Niq04,Niq06}.

Finally, in Section \ref{sec:concl} we conclude and outline the further work to be done.

\section{Inductive and Coinductive Types in Coq}\label{sec:indcoind}

In this section, we will give a short exposition of how inductive and coinductive types are defined and used in Coq. 
We will introduce several running examples.
The related work of developing the theory of corecursive definitions was done in  HOL and mechanised using Isabelle \cite{Paul97}. 
 For a more detailed introduction to Coq, see \cite{BC04}.

As we have already mentioned in the introduction, inductive data types are defined by introducing 
a few basic constructors that generate the elements of the new type.
\begin{definition}\label{ex:nat}
The definition of the inductive type of  natural numbers is built using two constructors \texttt{O} and \texttt{S}:
\begin{verbatim}
Inductive nat : Set :=
  | O : nat
  | S : nat -> nat.
\end{verbatim} 
\end{definition}

After the inductive type is defined, one can define its inhabitants
and functions on it.  Most functions defined on the inductive type must
be defined recursively, that is, by describing values for
different patterns of the constructors and by allowing calls to the
same function on variables taken from the patterns.  Thus, division by two on natural numbers is
computed by a recursive function \texttt{div2}:
$$ \left\{ \begin{array}{ll} \textrm{div2} (0) = 0 & \\
\textrm{div2}(\textrm{S}\  0) = 0 & \\
\textrm{div2}(\textrm{S} (\textrm{S}\  n')) = \textrm{S} (\textrm{div2}\  n').
\end{array}
\right.$$
And this can be modelled in Coq as follows.

\begin{definition}\label{ex:div2}
\begin{verbatim}

Fixpoint div2 n : nat :=
  match n with
  | O => 0
  | S O => 0
  | S (S n') => S (div2 n')
  end.
\end{verbatim}
\end{definition}

It is essential that recursive functions are defined over arguments of inductive types. For instance, natural numbers are given as arguments in the above definition. 

We can use the inductive type of natural numbers and the defined function \texttt{div2} to obtain a new recursive function.  The function of discrete logarithm of base 2 for natural numbers
is computed by a recursive function \texttt{log} satisfying the equation 
$$ \left\{ \begin{array}{ll} \textrm{log} (\textrm{S 0}) = \textrm{0} & \\
\textrm{log}(\textrm{S} (\textrm{S}\  n)) = \textrm{S} (\textrm{log}\  \textrm{S}(\textrm{div2}\ n)).
\end{array}
\right.$$
But this, as we further explain in Section \ref{sec:strguard}, cannot be modelled directly in Coq.
However, we can define the inductive predicate characterising the arguments for which the discrete logarithm is well-defined:

\begin{definition}\label{ex:log}
\begin{verbatim}

Inductive log_domain : nat -> Prop :=
| log_domain_1 : log_domain 1
| log_domain_2 : 
 forall p: nat, log_domain (S (div2 p)) -> log_domain (S (S p)).
\end{verbatim}
\end{definition}  

 It was observed in \cite{JR97}, that induction gives rise to initial algebras, while coinduction gives rise to final coalgebras; and the basic duality between algebras and coalgebras can be 
expressed as \emph{construction} versus \emph{observation}. Let us have a closer look at how this idea is realised in Coq.

The following is the definition of a coinductive type of infinite streams, built using one constructor \texttt{SCons}.
\begin{definition}\label{ex:str}
The type of streams is given by
\begin{verbatim}
CoInductive str (A:Set) : Set := SCons: A -> str A ->  str A.
\end{verbatim}
\end{definition}

Typically, a stream has the form $\texttt{SCons a s}$, where a is an element
 of some set $A$, and s is a stream. There exists a common convention to write \texttt{a :: s} for \texttt{SCons a s}.
Although syntactically the above definition is very similar to the definitions of the inductive types, this coinductive definition supports well the  dichotomy \emph{construction} 
versus \emph{observation}: given an infinite stream, 
we can only observe its head, and pass on to its tail. The tail will be infinite, too; and only its first element can be observed next.

\begin{definition}\label{ex:repeat}
The coinductive function \texttt{repeat} takes as argument  an element $a$ of some set $A$ 
and yields a stream where $a$ is repeated indefinitely:
\begin{verbatim}
CoFixpoint repeat (a: A): str A := SCons a (repeat a).
\end{verbatim}  
\end{definition}
Notably, we do not have to impose any type requirements on arguments of the function, but we require the produced values to be of coinductive type.

Properties of coinductive data often need to be expressed with coinductive predicates.
To prove some properties of infinite streams, we use the method of observation. 
For example, to prove that the two lists are \emph{bisimilar}, 
we must observe that their first elements are the same, and continue the process with the next. 

\begin{definition}\label{ex:bisimilar}
Bisimilarity is expressed in the definition of the following coinductive type:

\begin{verbatim}
CoInductive bisimilar_s:  str A -> str A -> Prop :=
|bisim: forall (a : A) (s s' : str A), bisimilar_s s s' -> 
                    bisimilar_s (SCons A a s)(SCons A a s').
\end{verbatim}
\end{definition}

The definition of \texttt{bismilar\_s} corresponds to the conventional notion of bisimilarity as given, e.g. in \cite{JR97}.
Lemmas and theorems analogous to the \emph{coinductive proof principle} of \cite{JR97} are proved in Coq and can be found in \cite{BC04}.

Infinite streams are not the only kind of data that is handled by coinductive machinery of Coq. We can work with different 
types of infinite data types, such as infinite binary trees or infinite expression trees, see also \cite{Acz00}. 
\emph{Expression trees} are trees in which every node has one or two children. The nodes of these trees are 
labelled with elements of sets $A$ and $B$, and we will call them $A$-nodes and $B$-nodes, respectively.
We will denote expression trees by $E(A,B)$.
The expression trees were extensively used in formalising real number arithmetic, see \cite{Niq06,EP97}.

\begin{definition}\label{ex:ETrees}
We coinductively define the expression trees in Coq: 
\begin{verbatim}
CoInductive ETrees (A B : Set) : Set :=
| A_node : A -> ETrees A B -> ETrees A B 
| B_node : B -> ETrees A B -> ETrees A B  -> ETrees A B.
\end{verbatim}
\end{definition}

We define a bisimilarity relation \texttt{bisimilar\_t} for this type in \cite{BertotKomendaCoq08}.
Also, in \cite{BertotKomendaCoq08} we show that 
 \texttt{bisimilar\_s} and \texttt{bisimilar\_t} 
are equivalence relations.

We have seen that in Coq, inductive types are domains of recursive functions and coinductive types 
are codomains of corecursive functions. We have also observed that syntactically, 
the definitions of inductive and coinductive types in Coq follow one and the same scheme.

\section{Termination and Productivity}\label{sec:termprod}
 In this section, we will discuss two computational concepts that 
depend on the recursive nature of \emph{inductive} and \emph{coinductive}
definitions,  
 and those are of \emph{termination} and  \emph{productivity}.

In Coq, as in any other type-theoretic theorem provers (HOL, PVS, and 
others, see \cite{BG01}), all computations must terminate. 
Because propositions are represented by types and proofs by programs, 
according to \emph{Curry-Howard} isomorphism \cite{Bar92,H80,Geu93},
we cannot allow non-terminating proofs, as they may lead to
inconsistency. There is a technical reason for the termination
requirement, too: to decide type-checking of dependent types, we need
to reduce type expressions to normal form (\cite{BG01} is a very good
survey of proof techniques used in type theory).

\begin{example}\label{ex:terminative}
The function \texttt{div2} from Definition \ref{ex:div2} is terminative. 

Given any natural number excluding $0$ as an input, the function \texttt{log}  
described in Section \ref{sec:indcoind} is terminative.  
\end{example}

For corecursive functions there is a dual notion to that of termination -  productivity.
The notion of termination is used to ensure totality of functions on \emph{finite} 
objects (initial algebras \cite{JR97}); while productivity is used to ensure 
totality of functions on \emph{infinite objects} (final coalgebras).
The infinite objects that we are going to use as running examples through 
the paper are streams and expression trees as defined in~\ref{ex:str}
and~\ref{ex:ETrees}.

The notion of productivity was first defined in \cite{Dij80,Sij89}, in terms of 
domain theory. For a very careful domain theoretic characterisation of 
productivity of streams and trees, see \cite{Niq04}.
However, we will omit the domain theoretic definitions here, and describe 
productivity from a computational point of view. Namely, we use recursive functions in order to
define classes of productive functions in Coq. We hope that this section will give the reader 
the opportunity to capture the spirit of a
functional approach  to productivity.
For more on productivity of infinite data structures, see, e.g. \cite{Buch05,Coq94,EGHIK07}.

Values in co-inductive types usually cannot be observed as a whole, 
because of their infiniteness.
Instead, they are often  described as some finite tree-like 
structures where some sub-terms still
remain to be computed and are described using unevaluated functions 
applied to arguments.  Values
in co-inductive types are said to be {\em productive} when all 
observations of fragments made using
recursive functions are guaranteed to be computable in finite time.

When the co-inductive type being considered is the type of 
streams, we can ask to
see the element of the stream at position \(n\) using the following function:
\begin{definition}
$$ \left\{ \begin{array}{ll} \texttt{nth 0}\  (\texttt{SCons a tl}) = \texttt{a} & \\
\texttt{nth}\  (\texttt{S n})\ (\texttt{SCons a tl}) = \texttt{nth n tl}
\end{array}
\right.$$
It is a regular structural recursive function with structural 
argument \(n\). 
\end{definition}

A given stream s
is productive if we can be sure that the computation of the list \texttt{nth n 
s} is guaranteed to terminate,
whatever the value of  \texttt{n} is (satisfying the condition for the function \texttt{nth} will be 
enough to ensure that it is
satisfied for any other recursive function, but this is hard to prove).

\begin{example}\label{ex:productive}
For any \texttt{n}, the value \texttt{repeat n}  is productive, (see also Definition \ref{ex:repeat}).
\end{example}

We can do the similar recursive observations on coinductive expression trees. We describe the observation of values at 
different positions in an expression
tree using a type of direction \texttt{direct} and a function \texttt{fetch} that takes 
sequences of directions to access
a given position in a tree.

\begin{definition}
Assuming the type \texttt{direct} is given by the following:\\
\texttt{Inductive direct : Type := L | R}, the function
\texttt{fetch} of the type\\ \texttt{forall A B:Set, list direct -> ETrees A B -> A+B} is defined as follows:

$$ \left\{ \begin{array}{ll} \texttt{fetch nil (A\_node a t)}\  = \  \texttt{inl a} & \\
\texttt{fetch (\_::tl) (A\_node a t)}\  = \  \texttt{fetch tl t}\\
\texttt{fetch nil (B\_node b t1 t2)}\  = \  \texttt{inr b}\\
\texttt{fetch (L::tl) (B\_node b t1 t2)}\  = \  \texttt{fetch tl t1}\\
\texttt{fetch (R::tl) (B\_node b t1 t2)}\  = \  \texttt{fetch tl t2}
\end{array}
\right.$$

\end{definition}

An expression tree \texttt{t} is said to be productive if the computation of  
\texttt{fetch l t} is guaranteed to terminate
whatever the value of \texttt{l} is.

We call a function \emph{productive at the input value $i$}, 
if it outputs a productive value at $i$. This understanding of 
productivity of functions differs slightly from 
the approach of \cite{Sij89,Niq04,Niq06}, where a function is said 
to be productive if it takes productive values as an input and outputs productive values. 
Let us explain this on the following examples.

The function \emph{dyn} defined below takes an arbitrary value as an input. It returns productive values only for some inputs.
\begin{definition}\label{df:dyn}
Let $A$, $B$ be of type \texttt{Set}. For a predicate $P :\  B \rightarrow \texttt{bool}$ and functions
$h :\  B \rightarrow A$, $g,\  g' :\  B \rightarrow B$, we define the function \emph{dyn} as follows:

$$ \textrm{dyn (x)} \  =  \left\{ \begin{array}{ll} \textrm{SCons}\  h(x)\  (\textrm{dyn}\ 
(g( x))) & \  \textrm{if}\  P(x) \\
\textrm{dyn}\  (g' (x)) & \textrm{otherwise.}
\end{array}
\right.$$
\end{definition}

Thus, unlike \cite{Niq04,Niq06}, we do not require productive functions to take coinductive values as an input. 
The only requirement we impose is that the produced data must be coinductive. 
This approach is consistent with the understanding of coinductive functions in Coq: 
arguments of corecursive functions can be of any type, and only the produced values are required to be of coinductive type.

There is a tradition of studying {\em productive} functions, probably 
meaning that these functions are \emph{totally productive} once given productive values as arguments.  
However, in this work we
want to study functions that are only \emph{partially productive}, that is, 
functions that will return
productive values only for a subset of their input type, a subset which 
we characterise precisely
using inductive and coinductive predicates.
We illustrate this further as follows.   

Consider the filter function  on streams that was formalised in \cite{Bert05} and was used to filter prime numbers.
\begin{definition}{\bf{(Filter for streams)}}.\label{ex:filter}\\ 
 For a given predicate $P$,
$$ \textrm{filter (SCons x tl)} =  \left\{ \begin{array}{ll} \textrm{SCons x (filter tl)} & \  \textrm{if}\  P(x) \\
\textrm{filter tl} & \textrm{otherwise.}
\end{array}
\right.$$
\end{definition}

The filter function examines the first element \texttt{x} of a given
list (\texttt{SCons x tl}) for a property $P$, and, in case the property is satisfied, it uses \texttt{x} to
form a new list. Then it recursively examines the tail of the stream.

In general, filter functions on streams make it possible to define 
non-productive values. The filter function above can be non-productive on certain values and for certain predicates.

\begin{example}\label{ex:omega0}
For instance,
 computing \texttt{nth 0 (filter even (repeat 1))} provokes the following 
computation:

\texttt{filter even (repeat 1) repeat 1} $\rightsquigarrow$ \texttt{filter even (1::repeat 1)} 
        $\rightsquigarrow$ \texttt{filter even (repeat 1)}...

The first arrow comes from computing a little portion of (repeat 1). The second 
arrow comes from observing
that 1 is not even and reducing the definition of filter. This leads to 
the same position as initially without
producing the first element of the stream required by (\texttt{nth 0}). The same 
computation should be triggered again
and indefinitely.
\end{example}

The method of formalising corecursive functions that we propose in this paper, makes it possible to formalise 
such functions in Coq, using inductive and coinductive predicates to characterise the arguments on which these functions output productive values.
By abuse of terminology that causes no confusion,  we will call these functions ``productive''. 

We conclude this section with another important example of a partially productive corecursive function.
The following functions generalise the filter from Definition \ref{ex:filter} to the case of expression trees with dynamic filtering:
\begin{definition}{\bf{(Dynamic Filter for Expression Trees)}}.\label{ex:efilter}\\
Let $P$ and $P_1$ be predicates, and let 
$h_1 : A_1 \rightarrow A_2$, $h_2: B_1 \rightarrow  A_2$, 
$h_1': A_1 \rightarrow A_1$, $h_2' : B_1 \rightarrow B_1$, 
$h3: \texttt{ETrees} \ A_1 B_1 \rightarrow \texttt{ETrees}\  A_1 B_1$. The latter functions will
 ``dynamically'' change the filtered values in the process of filtering: 

$$ \textrm{e\_filter} \  (\textrm{A\_node}\  a\  t_1) =  \left\{ \begin{array}{ll} \textrm{A\_node}\  h_1(a)\  (\textrm{e\_filter}\ 
(\textrm{A\_node}\  h_1'(a)\  t_1)) & \  \textrm{if}\  P(a) \\
\textrm{e\_filter}\  h_3 \ (\textrm{A\_node}\ a\ t_1) & \textrm{otherwise.}
\end{array}
\right.$$

$$ \textrm{e\_filter} \  (\textrm{B\_node}\  b\  t_1 \ t_2) =  \left\{ \begin{array}{ll} \textrm{A\_node}\  h_2(b)\  (\textrm{e\_filter}\ 
(\textrm{B\_node}\  h_2'(b)\  t_1 \ t_2)) & \  \textrm{if}\  P_1(b) \\
\textrm{e\_filter}\  h_3 \ (\textrm{B\_node}\ b\ t_1 \ t_2) & \textrm{otherwise.}
\end{array}
\right.$$
\end{definition}

The function was used to establish a normalisation algorithm for
an admissible representation of a closed interval of real numbers
in \cite{Geu93,Niq04}.

We have discussed the reasons why termination and productivity are
important for our theory, and also we showed how they relate to each
other. We introduced  important examples of partially productive functions.
In the next section, we will discuss the way how terminative and productive functions are syntactically defined in Coq.

\section{Structural Recursion and Guardedness; Method of Ad-hoc Predicates}\label{sec:strguard}

There are two syntactic tests that ensure termination and productivity of functions in Coq, they are called 
\emph{structurally smaller calls condition} \cite{Coq92,PM93} and
\emph{guarded-by-constructors condition} \cite{Gim96}. 

A structurally recursive definition is such that every recursive call
is performed on a structurally smaller argument. In this way we can be
sure that the recursion terminates.

Guardedness is a sufficient condition for productivity.
It is described in two steps.  The first step defines 
\emph{pre-guarded positions}.
A position is pre-guarded if it occurs as the root of the function body, 
or if it is a direct
sub-term of a pattern-matching construct or a conditional statement, 
which is itself in
a pre-guarded position.  

The second step defines \emph{guarded positions}.  A
position is guarded
if it occurs as a direct sub-term of a constructor for the co-inductive 
type that is being defined
and if this constructor occurs in a pre-guarded position or a guarded 
position.  A corecursive function is
\emph{guarded} if all its corecursive calls occur in guarded 
positions.

Guardedness ensures that at least one constructor of the co-inductive 
type is produced for each
corecursive call, and thus at least a fragment of corecursive data is 
produced each time a
corecursive call occurs.

\begin{example}
The function \texttt{div2} from Definition \ref{ex:div2} is structurally recursive. 

The function \texttt{repeat} from Definition \ref{ex:repeat} is guarded.
\end{example}

Definitions where the recursive calls are not required to be on structurally smaller arguments, that is, 
where the recursive calls can be performed on any argument, are called \emph{general recursive}.
Many important algorithms, such as the algorithm of computing logarithm discussed in Section \ref{sec:indcoind}, 
are not structurally recursive but general recursive.

The standard way of handling general recursion in constructive type theory uses a well-founded recursion principle derived 
from the accessibility predicate \texttt{Acc} \cite{Acz77,Nord88}.
The idea behind the accessibility predicate is that an element $a$ is \emph{accessible} by a 
relation $<$ if there is no infinite decreasing sequence starting from $a$.
A set $A$ is said to be \emph{well-founded} with respect to $<$ if all its elements are 
accessible by $<$. Hence, to guarantee that a general recursive algorithm  that performs 
the recursive calls on elements of type $A$ terminates, we have to prove that $A$ is well-founded and 
that the arguments supplied to the recursive calls are smaller than the input.

In Coq, the method was implemented in \cite{Hue88,BB00}, see also \cite{BC04}. 
The method of using accessibility predicates was improved by Bove in her thesis \cite{Bove02phd} 
and series of papers \cite{Bov01,BoveC01,Bove02,BoveC05}.  The core of the improvement proposed 
by Bove was to separate computational and logical parts of the definitions of general recursive 
algorithms. That is, the method amounts to defining an inductive special-purpose accessibility
 (ad-hoc) predicate that characterises the inputs on which the algorithm terminates. 
Proving that a certain function is total amounts to proving that the corresponding accessibility 
predicate is satisfied on every  input.  

\begin{example}\label{ex:logadhoc}
We continue Definition \ref{ex:log}, and formalise the function log defined in Section \ref{sec:indcoind}.
We need an inversion lemma about the predicate \texttt{log\_domain} expressing 
that when \texttt{x} has the form ``\texttt{S (S p)}", if \texttt{x} is in the domain, 
then ``\texttt{S (div2 p)}" also is:

\begin{verbatim}
Lemma log_domain_inv :
forall x p : nat, log_domain x -> x = S(S p) -> 
                  log_domain (S (div2 p)).
\end{verbatim}

At each recursive call of the function, we need to use the above
\emph{inversion lemma} stating that the proof argument for recursive
call can be deduced from the initial proof argument.

We also need to express that $0$ is not in the domain of the function:\\
\texttt{Lemma log\_domain\_non\_0: forall x :nat,}
\texttt{log\_domain x -> x} $\neq$ \texttt{0.}

Now we can use the ad-hoc predicate \texttt{log\_domain\_inv}, together with the function \texttt{div2} 
from Definition \ref{ex:div2} and
define the function \texttt{log} as a usual structurally recursive
function, but where the structural argument is the {\em proof
  argument}; we box the proof arguments in the example below.

\begin{alltt}
Fixpoint log (x:nat)\framebox{(h: log_domain x)}{struct h} : nat :=
match x as y return x = y -> nat with
| 0 => fun h' => False_rec nat \framebox{(log_domain_non_0 x h h')}
| S 0 => fun h' => 0
| S (S p) => 
   fun h' => S (log (S (div2 p)) \framebox{(log_domain_inv x p h h'))}
end (refl_equal x).
\end{alltt}
It is important that the Coq checker can recognise \texttt{log\_domain\_inv  x p h h'} as a structurally 
smaller proof with respect to \texttt{h}. See also \cite{BC04}, Section 15.4. 
 \end{example}

However, the source of our interest in this method of ad-hoc predicates lies not in general recursive functions, but
 in productive non-guarded corecursive functions.
In the same way as the syntactic
 structurally recursive condition removes 
many useful terminative functions from the picture, the syntactic
guardedness condition of Coq rejects many productive functions.

Informally speaking, the guardedness condition insures that 
\begin{itemize}
\item[*]\label{it:a} each corecursive call is made under at least one constructor;
\item[**]\label{it:b} if the recursive call is under a constructor, it does not appear as an argument of any function.
\end{itemize}

Violation of any of these two conditions makes a function rejected by the guardedness test in Coq.

\begin{example}
The corecursive functions from Definitions \ref{ex:filter}, \ref{ex:efilter} and \ref{df:dyn} 
are not directly formalisable in Coq, because they do not satisfy the guardedness condition *.
However, we know about many useful examples of filters from Definitions \ref{ex:filter} and 
\ref{ex:efilter} that are productive, \cite{Bert05,Niq06,EP97}.
\end{example}

All the examples we  have given so far exhibit \emph{partially} productive non-guarded functions. And this may give the reader 
an impression that it is ``partiality" of corecursive functions that makes them non-guarded. However, totally productive functions can be non-guarded, too.
Moreover, we claim that every terminative function gives rise to a
      non-guarded \emph{totally} productive function, or, in other words, 
one can always craft a totally productive non-guarded function using a terminative function.

\begin{example}
For example, terminative function $ x - 1$ gives rise to the totally productive
function \texttt{f: str nat -> str nat:}\\
$$ \texttt{f (x::y::tl)} =  \left\{ \begin{array}{ll}  \texttt{x::f(y::tl)} & \  \textrm{if}\  x \leq y  \\
\texttt{f((x-1)::y::tl)} & \textrm{otherwise.}
\end{array}
\right.$$ 
This second corecursive call does not satisfy the guardedness condition *.
\end{example}

In general, the fact that one uses a terminative function to define a corecursive function guarantees 
productiveness, but in the same time one can always craft the resulting corecursive function in such a 
way that a corecursive call will not be under a constructor, thus violating guardedness condition *.

The next example illustrates the class of functions that fail to satisfy the guardedness condition **.

\begin{example}
Consider the following function computing lists of ordered natural numbers:
\begin{verbatim}
nats = (SCons 1  (map (+ 1) nats)).
\end{verbatim}
where the function \texttt{map} above is defined as follows:
\begin{verbatim}
map f (s: str): str := Cons (f (hd s)) (map f (tl s)).
\end{verbatim}
That is, we start with the list with 1, then add (\texttt{+ 1}) to the head of the list to get the second element 2; 
and continue the same computation on the tail.
 
The recursive call here is made under the constructor \texttt{SConc}, but it also appears on the argument place of the function \texttt{map}. 
The latter fact violates the guardedness condition **, and hence the function will be rejected by Coq.
Despite of this, the value \texttt{nats} is known to be productive.

The similar problem arises with the Example 1 in \cite{EGHIK07}: the function is productive but non-guarded.
\end{example}

In \cite{Bert05} was found a Coq formalisation of a coinductive \emph{filter} function from Definition \ref{ex:filter}.
The work was aimed at showing that values
produced through filter functions could still be described as guarded 
corecursive functions.
Notably, the solution involved the method of building an ad-hoc predicate similar to the ad-hoc predicate of Bove.
The fact that the method introduced for inductive algorithms is expandable to coinductive ones is significant in 
its own right. But the example of \cite{Bert05} exhibited even more:
the method of formalising corecursive filter functions requires separation of not only ``logical'' and 
``computational'' parts of algorithms, but also of inductive and coinductive parts.

In the remaining sections, we develop the general method of formalising productive functions that fail to satisfy 
the guardedness condition *. Our method does not cover the class of productive functions that do not satisfy the guardedness condition **. 
Such functions were studied (e.g.) in \cite{Abel06phd,EGHIK07}.

\section{Inductive Components  of Corecursive  Functions}\label{sec:coadhoc}

For each productive function, we will describe two modalities,
\texttt{eventually} and \texttt{infinite} that we are going to use when
characterising the inductive and coinductive components of productive functions. 
These are variations of $\Diamond$  and $\Box$ specified 
coalgebraically in \cite{Jac02}, and they originate from the temporal modalities introduced in \cite{Pnu81}.

The conventional definitions of \texttt{eventually} and \texttt{infinite} as formalised, (e.g.) in \cite{BC04}, 
express whether a given stream \texttt{s} satisfies some 
given property \texttt{P} at least once or infinitely many times.
We modify these predicates in order to characterise the conditions for
a corecursive function to perform a guarded corecursive step. In this section, we consider \texttt{eventually} 
and its role in defining a recursive function 
which characterises the \emph{inductive component} of a given corecursive function. 

We propose to take as a starting point the defining equation of the corecursive function we wish to formalise; 
as e.g., Definitions \ref{ex:filter}, \ref{ex:efilter}, \ref{df:dyn}.  
Then we define predicates
on the input types of these functions.  The predicate {\tt 
eventually} captures the conditions
for the corecursive function to perform the next guarded corecursive step.
For any given function, and on any given data type, we can define \texttt{eventually} systematically as follows:

{\textbf{1.}} \emph{The predicate \texttt{eventually} is defined inductively, with one constructor for 
each branch appearing in the function definition.}

For example, we will notice that in Definitions \ref{ex:evstr} and \ref{df:evetree} 
the number of constructors we use to define \texttt{eventually} will vary from two to four, according to the number of 
branches appearing in Definitions \ref{ex:filter}, \ref{ex:efilter}.    
 
{\textbf{2.}} \emph{When a branch contains only
guarded recursive calls, the constructor expresses that 
the input data satisfies the \texttt{eventually}
predicate as soon as it satisfies all the conditions needed to reach this branch.}

For instance, the {\tt dyn} function from Definition \ref{df:dyn} performs
a boolean test on the predicate {\tt P}, and returns the
value {\tt SCons (h x) (dyn (g x))} if {\tt (P x)} is {\tt true}, or
the value {\tt (dyn (g' x))} if {\tt (P x)} is {\tt false}.
Each of these values contains a recursive call, but the first call
on {\tt (g x)} is guarded while the other call on {\tt (g' x)} is
non-guarded.

For the first recursive call, we can directly provide a constructor to
this inductive predicate that states that {\tt x} satisfies the predicate
if {\tt (P x)} is {\tt true}. (We choose to name the {\tt eventually} predicate {\tt eventually\_dyn}.)
\begin{verbatim}
 ev_dyn1 : P x = true  -> eventually_dyn x
\end{verbatim}

{\textbf{3.}} \emph{When a branch contains non-guarded recursive calls, the constructor
expresses  that the input data satisfies
the predicate as soon as it satisfies all the conditions leading to this 
branch, and the inputs to all non-guarded recursive calls satisfy the
predicate.}

Let us return to the \texttt{dyn} function. For the second recursive call, 
we provide a constructor stating that {\tt x}
satisfies the predicate when {\tt (P x)} is {\tt false} only if the
recursive call would then reach a value that already satisfies the
predicate:
\begin{verbatim}
 ev_dyn2 : P x = false  -> eventually_dyn (g' x) -> eventually_dyn x
\end{verbatim}

We have covered all possible recursive branches in the behaviour of {\tt dyn},
so we can collect the constructors in an inductive definition:
\begin{definition}
\begin{verbatim}

Inductive eventually_dyn (x:B) : Prop :=
| ev_dyn1 : P x = true  -> eventually_dyn x
| ev_dyn2 : P x = false  -> eventually_dyn (g' x) -> eventually_dyn x.
\end{verbatim}
\end{definition}

The same systematic approach gives the predicates {\tt eventually\_s}
for the filter on streams and {\tt eventually\_t} for
the filter on expression trees:
\begin{definition}\label{ex:evstr}
\begin{verbatim}

Inductive eventually_s: str A ->  Prop :=
| ev_b: forall x s, P x ->  eventually_s (SCons A x s)
| ev_r: forall x s, 
  ~ P x -> eventually_s s -> eventually_s (SCons A x s).
\end{verbatim}
\end{definition}

Note that \texttt{eventually\_t} has four constructors, depending on whether the 
input data is an A-node or a B-node and depending on whether 
the observable data carried by this node satisfies a predicate P or not, 
precisely as in Definition \ref{ex:efilter}:
\begin{definition}\label{df:evetree}
\begin{verbatim}
Inductive eventually_t: ETrees A1 B1  -> Prop :=
|ev_rB: forall x t t1 ,  
  ~P1 x -> eventually_t (h3 (B_node A1 B1 x t t1))  -> 
  eventually_t (B_node A1 B1 x t t1) 
|ev_bB: forall x t t1, P1 x  -> eventually_t (B_node A1 B1 x t t1)
|ev_rA: forall x t,  
  ~P x -> eventually_t (h3 (A_node A1 B1 x t)) ->  
  eventually_t (A_node A1 B1 x t) 
|ev_bA: forall x t, P x -> eventually_t (A_node A1 B1 x t).
\end{verbatim}
\end{definition}

Thus, \texttt{eventually\_dyn}, \texttt{eventually\_t} and \texttt{eventually\_s} 
are constructed using one systematic method.

The \texttt{eventually} predicates are satisfied if the conditions for
the recursive function to perform a guarded corecursive call are satisfied
at least once.  In the future, we will say that \texttt{eventually} ensures
only the
\emph{first-step-productivity}, as opposed to the conventional
\emph{productivity} of Section \ref{sec:termprod}.

Using the inductive definition of \texttt{eventually} predicates,
we can describe the first component of the function we want to compute.
This component is a recursive function that performs all the computations
and tests that lead to the first guarded corecursive call.
For lack of a better name, we will call this function the
{\em inductive component} of the corecursive function being defined.
Separating this inductive component from the rest of the function
behaviour is interesting: it will hide all the non-guarded corecursive calls
and make it possible to go directly from one guarded call to the next one.

This inductive component is modelled by following the technique of ad-hoc
predicates where the chosen predicate is the {\tt eventually} predicate.

The method of ad-hoc predicates relies on inversion lemmas.  
Each inversion lemma 
expresses that the argument to the recursive call satisfies the 
{\tt eventually} predicate if the initial argument does.  Moreover, the
proof of each inversion lemma is carefully crafted to make sure
that the lemma's conclusion is understood as a {\em structurally smaller
proof} with respect to one of the lemma's premises (this is the essence
of recursion on an ad-hoc predicate, see \cite{BC04}).

There is one inversion lemma needed for the {\tt dyn} function and the filter on streams. Two
inversion lemmas are needed to formalise the filter on trees. In general, the number of the required 
inversion lemmas equals to the number of non-guarded calls in the initial corecursive function.
\begin{lemma}[Inversion Lemmas]
\begin{verbatim}
Lemma eventually_dyn_inv :
 forall x, eventually_dyn x -> P x = false -> eventually_dyn (g' x).

Lemma eventually_s_inv: forall (s : str A), 
 eventually_s s -> forall x s', s = SCons A x s' -> 
                                           ~ P x ->  eventually_s s'.
 
Lemma eventually_t_inv:
 forall t : ETrees A1 B1, eventually_t t -> 
 forall (x: A1) (t': ETrees A1 B1), t =  (A_node A1 B1 x t') ->
                     ~ P x -> eventually_t (h3 (A_node A1 B1 x t')).

Lemma eventually_t_inv':
 forall t : ETrees A1 B1,  eventually_t t ->  
 forall(x:B1) (t' t1: ETrees A1 B1), t = (B_node A1 B1 x t' t1) ->
                  ~P1 x -> eventually_t (h3 (B_node A1 B1 x t' t1)).
\end{verbatim}
\end{lemma}
We give the full proofs of these inversion lemmas in
\cite{BertotKomendaCoq08}.

We use these inversion lemmas to define the inductive component of
corecursive functions: this inductive component does all computations
until it performs the first guarded corecursive call.  For instance, for
the function on streams, the inductive component performs all computation
steps until one reaches the first element in the stream that satisfies
the property {\tt P} used for filtering.

The data produced by the inductive component is organised in
two parts: the first part contains the observable data that is included
in the head constructor of the output.  Such a head constructor is necessarily
present, because the {\tt eventually} predicate holds and this
means that there is a guarded corecursive call.  The second part contains
the argument for the next recursive call of the initial corecursive function.

The inductive component can perform recursive calls to itself, but they
go from values satisfying the {\tt eventually} predicate to values satisfying
the {\tt eventually} predicate, thanks to the inversion lemmas.

\begin{definition} The inductive component of the recursive function for {\tt dyn} 
is called {\tt pre\_dyn}; the additional (proof) arguments are boxed below:
\begin{alltt}
Fixpoint pre_dyn (x:B)\framebox{(d:eventually_dyn x)} \{struct d\} : A*B :=
  match P x as b return P x = b -> A*B with
    true => fun t => (h x, g x)
  | false => fun t => pre_dyn (g' x) \framebox{(eventually_dyn_inv x d t)}
  end (refl_equal (P x)).
\end{alltt}
This code exhibits two aspects of programming with ad-hoc predicates, which
 make them complicated to read for neophytes: first, the function has
an extra proof argument and recursive calls must also have this extra
proof argument.  Second, the body of the function contains a pattern-matching
construction that must be applied to a proof of {\tt P x = P x}.  Inside
the proof matching construct, this equality is transformed differently
in the two branches: in one branch it takes the form of {\tt P x = true}
while in the other it takes of the form of {\tt P x = false}.
\end{definition}

For the function {\tt e\_filter}, the inductive component is
named {\tt pre\_filter\_t}.
\begin{definition} 
\begin{verbatim}

Fixpoint pre_filter_t (t: ETrees A1 B1)
(h: eventually_t t){struct h} : A2 * ETrees A1 B1  :=
match t as b return t = b ->  A2*ETrees A1 B1  with
|A_node x t' => fun heq : t = (A_node A1 B1 x t') => 
  match P_dec x with
  |left _ => (h1 (inl B1 x), A_node (fst (h2 (inl B1 x))) t')
  |right hn => pre_filter_t (h3 (A_node A1 B1 x t'))
                         (eventually_t_inv t h x t' heq hn)
  end
|B_node x t' t1 => fun heq': t = (@B_node A1 B1 x t' t1) =>
  match P1_dec x with
  |left _ => (h1 (inr A1 x), B_node (snd (h2 (inr A1 x))) t' t1)
  |right hn' => pre_filter_t(h3 (B_node A1 B1 x t' t1))
                       (eventually_t_inv' t h x t' t1 heq' hn')
  end
end (refl_equal t).
\end{verbatim}
\end{definition}

We formulate the omitted inductive component of the filter for streams in \cite{BertotKomendaCoq08}.

In this section, we have shown the algorithm to formalise the
inductive components of several recursive functions.
It remains to describe how computation can carry on beyond the first
guarded corecursive call.

\section{Coinductive Components of  Corecursive Functions}\label{sec:coind}
  
In this section we characterise the coinductive part of corecursive functions.
Corecursive computations are introduced by repeating computations performed
by the inductive component defined in the previous section.  Now, because
the inductive component finds guarded corecursive call, 
guarded co-recursion can take place.  However, this can only happen if
recursive calls satisfy the \texttt{eventually} predicate repeatedly.
We need an extra predicate to express this.

We define a coinductive predicate \texttt{infinite}, a variation of which was denoted by $\Box$
in the examples of coalgebraic specifications by Jacobs \cite{Jac02}.
It is notable that M.~Niqui, who suggested a similar formalisation of filter
functions for streams, used a predicate $\texttt{is\_infinite?}$ in order to extract the
desired guarded filter, see \cite{Niq04}. 

We suggest to define {\tt infinite} predicates co-inductively by expressing
that a value satisfies this predicate if it satisfies the {\tt eventually}
predicate and the second part computed by the inductive component satisfies
the {\tt infinite} predicate again.  Remember that this second part is
the argument for the next recursive call of the main corecursive function.
\begin{definition}
 For the {\tt dyn} function,
 we choose to call the {\tt infinite}
predicate {\tt infinite\_dyn}.  It is described as follows:
\begin{verbatim}
CoInductive infinite_dyn (x : B): Prop :=
  di : forall d: eventually_dyn x,
   infinite_dyn (snd (pre_dyn x d)) -> infinite_dyn x.
\end{verbatim}
For the {\tt filter\_s} function,  we choose to call the predicate
\texttt{infinite\_s}:
\begin{verbatim}
CoInductive infinite_s : str ->  Prop :=
  al_cons: forall (s:str A) (h: eventually s),
   infinite_s (snd(pre_filter_s s h)) ->  infinite_s s.
\end{verbatim}
For the {\tt e\_filter} function we choose to call the predicate
\texttt{infinite\_t}:
\begin{verbatim}
CoInductive infinite_t : ETrees A1 B1 -> Prop :=
  cf: forall  (t:ETrees A1 B1) (h:eventually_t t), 
    infinite_t (snd (pre_filter_t t h)) -> infinite_t t.
\end{verbatim}
\end{definition}
Notice that the shape of \texttt{infinite} predicates for all these functions
is almost identical.

The {\tt infinite} predicate describes exactly those arguments to the
function for which the function is guaranteed to be productive.  Thus, we
will reproduce the scheme already followed in recursion with respect to an
ad-hoc predicate: the recursive function will be modelled in Coq by a
function that receives an extra argument, a proof that the initial argument
satisfies the {\tt infinite} predicate.
Computation will be performed by repeatedly calling the inductive component
of the function.
Since the inductive component function can only compute on arguments that
satisfy the {\tt eventually} predicate, we need a lemma stating that
the {\tt infinite} predicate implies the {\tt eventually} predicate.  This
lemma is always obtained using a very simple proof by pattern matching.

For the {\tt dyn} and \texttt{filter\_t} functions the lemma has the following statements:
\begin{lemma}
\begin{verbatim}

Lemma infinite_eventually_dyn : 
  forall x, infinite_dyn x -> eventually_dyn x.
\end{verbatim}
\begin{verbatim}
Lemma infinite_eventually_t :
  forall t, infinite_t t -> eventually_t t.
\end{verbatim}
\end{lemma}

After the computation of the first component, computation must go on, with
recursive calls obtained from the second part of the data computed in
the inductive component.  However, this data must also be associated with
a proof that it satisfies the {\tt infinite} predicate.  This is expressed
by means of the following lemmas, given for \texttt{dyn} and \texttt{e\_filter}:
\begin{lemma}

\begin{verbatim}

Lemma infinite_always_dyn : 
  forall x, infinite_dyn x -> forall e: eventually_dyn x,
  infinite_dyn (snd (pre_dyn x e)).
\end{verbatim}

\begin{verbatim}
Lemma infinite_always_t : forall(t : ETrees A1 B1)(h:infinite_t t),
     infinite_t (snd (pre_filter_t t (infinite_eventually_t t h))).  
\end{verbatim}
\end{lemma}

We give the proofs of these lemmas in \cite{BertotKomendaCoq08}.

Finally, we use these two categories of lemmas to formalise the main 
functions from Definitions \ref{df:dyn}, \ref{ex:efilter}
as guarded corecursive functions.

\begin{definition}\label{df:filter}
For the {\tt dyn} function the definition has the following shape (proof arguments are boxed):
\begin{alltt}
CoFixpoint dyn (x:B)\framebox{(h:infinite_dyn x)} : str :=
SCons (fst (pre_dyn x \framebox{(infinite_eventually_dyn ev x h)}))
 (dyn _ \framebox{(infinite_always_dyn x h (infinite_eventually_dyn x h))}).
\end{alltt}
For the {\tt e\_filter} function, the definition has the following shape:
\begin{verbatim}
CoFixpoint e_filter (t:ETrees A1 B1)(h: infinite_t t)
  : ETrees A2 B2 :=
match t return infinite_t t -> ETrees A2 B2 with
| A_node x t' => fun h' : infinite_t (A_node x t') =>
    A_node (fst (pre_filter_t t (infinite_eventually_t h'))) 
           (e_filter _ (infinite_always h'))
| B_node x t' t2 => fun h' : infinite_t (B_node x t' t2) =>
    A_node (fst (pre_filter_t t (infinite_eventually_t h'))) 
           (e_filter _ (infinite_always h'))
  end.
\end{verbatim}
\end{definition}
The guardedness of the functions is simply achieved by the presence of
constructors {\tt SCons} or {\tt A\_node}.  The recursive calls receive
the extra proof arguments.  Notice that we don't need to give explicitly
the value on which the recursive call is performed, because this value can
be inferred from the second proof argument.  This explains why we used
the anonymous value place holder {\tt \_} in several places.

We formalise the filter from Definition \ref{ex:filter} in \cite{BertotKomendaCoq08}.

In this Section, we have given a systematic characterisation of the 
coinductive components of productive functions.
We employed the predicate \texttt{infinite} to obtain the guarded 
formalisation of productive functions in Coq.  It remains to provide tools to reason on the
functions we modelled.

\section{Proving Properties About the Models}\label{sec:eqlemma}
In this section, we give the proofs of \emph{recursive equation lemmas}, 
statements that express that the functions we obtain
in Definition~\ref{df:filter} really are good models
of the non-guarded functions from Definitions
\ref{df:dyn}, \ref{ex:filter}, \ref{ex:efilter}.  
We discuss how such proofs can be obtained in a systematic way, outlining
a few useful lemmas about these functions.

The inductive component and coinductive component for each corecursive
function are modelled by Coq functions which take both a regular
argument (named {\tt x} in the function {\tt dyn} and {\tt t} in the
function {\tt e\_filter}) and a proof about this argument (named {\tt h}
in both {\tt dyn} and {\tt e\_filter}).  Of course, we expect the
result to depend on the regular argument, but we do not expect the
computation to depend on the value of the proof.  We only expect
the computation to depend on the 
existence of this proof.  In other words, when receiving a given argument 
{\tt x} and two different proofs {\tt e1} and {\tt e2}
that {\tt x} satisfies the {\tt infinite\_dyn} predicate, we expect
the resulting values {\tt dyn x e1} and {\tt dyn x e2} to be the same.
This needs to be expressed in lemmas which we call {\em
  proof-irrelevance} lemmas.
\begin{lemma}[Proof-irrelevance Lemmas for \texttt{dyn}].\\
For the {\tt dyn} function the proof irrelevance lemmas have the
following shape:
\begin{verbatim}
Lemma pre_dyn_prf_irrelevant :
 forall x e1 e2, pre_dyn x e1 = pre_dyn x e2.

Lemma dyn_prf_irrelevant :
  forall x x' (i: infinite_dyn x) (i':infinite_dyn x'), x = x' ->
  bisimilar_s (dyn x i) (dyn x' i').
\end{verbatim}
\end{lemma}
The proofs of these lemmas are given in \cite{BertotKomendaCoq08}.
Also, in \cite{BertotKomendaCoq08} we formulate and prove the similar proof irrelevance lemmas 
for filters on streams and expression trees.
A similar proof irrelevance lemma was used in \cite{Niq06}; a very good study of applications of the 
Principle of Proof Irrelevance in type theory can be found in \cite{Barthe98}.

Note that the above  proof irrelevance lemma, as well as analogous lemmas for 
filters on streams and trees \cite{BertotKomendaCoq08}, 
uses the ``bisimilar'' relation from Definition \ref{ex:bisimilar} instead of the
equality relation.  

In coalgebra, one has the \emph{coinductive proof principle} 
which states that bisimilar objects are equivalent, \cite{JR97}.
The constructive theory distinguishes equality and bisimilarity; and in many cases we can 
only give constructive proofs of bisimilarity of infinite objects; but not of their equality.

Nevertheless, for all practical
purposes, bisimilar objects can be used like equal objects.  
For example, one can prove in Coq that equality implies bisimilarity; 
or that for finite objects bisimilarity implies equality. 
Lemmas and theorems relating bisimilarity and equality can be found in \cite{BC04}.

It remains to show that the functions \texttt{dyn}, \texttt{filter} 
and \texttt{e\_filter} modelled in Coq actually perform the
computation that was initially intended.  The main idea is to perform as
in \cite{BB00,BalaaBertot02} and to prove a recursive equation.
However, we have to cope with the last difficulty.  The models we
obtain have an extra argument, which is used to restrict the function to
the inputs on which they really are productive.

We circumvent the difficulty by integrating extra proof arguments
in the formulation of the recursive equation.  As a consequence, we need
a few more lemmas expressing that all recursive calls that happen in
the recursive equation are made on values that satisfy the {\tt infinite}
predicate, as soon as the initial argument already satisfies this predicate.
These lemmas will be called {\em step-lemmas}.
\begin{example}
For instance, the specification of {\tt dyn} imposes that there
should be a recursive call {\tt dyn (g' x)} as soon as {\tt P x = false}.
Thus, we need to have a lemma that states that {\tt (g' x)} satisfies
{\tt infinite\_dyn} as soon as {\tt x} does and {\tt P x = false}.
\end{example}

\begin{lemma}[Step-lemmas for \texttt{dyn}].\\
For the function {\tt dyn}, there are two step lemmas:
\begin{verbatim}
Lemma dyn_step1 :
  forall x, P x = true -> infinite_dyn x -> infinite_dyn (g x).

Lemma dyn_step2 :
  forall x, P x = false -> infinite_dyn x -> infinite_dyn (g' x).
\end{verbatim}
\end{lemma}
We formalise the proofs for these lemmas in \cite{BertotKomendaCoq08}.
The first lemma expresses that the recursive call on {\tt g x} is
legitimate when {\tt P x} is true, while the second lemma expresses
that the recursive call on {\tt g' x} is legitimate when {\tt P x} is false.

We prove the similar step-lemmas for filters on expression trees and streams in 
\cite{BertotKomendaCoq08}.
They are all formulated uniformly with the step-lemmas above. For expression 
trees, for example, we need four step lemmas, to give an account both  for the 
cases when \texttt{P x} is true or false, and for the two constructors used to 
define expression trees. Conceptually, all these lemmas can be formulated in a uniform way.

We can then formulate the recursive equation for the functions we modelled.
This recursive equation is not as easy to read as the initial description,
because proof arguments have been added in many places, but if we ignore
these proof arguments, the initial intent appears clearly.

For the {\tt dyn} function, the equation lemma has the following shape,
where bisimilarity is again used instead of equality.
To help the reader to distinguish proof content from algorithmic content,
we boxed the proof arguments.
\begin{theorem}[Recursive Equation Lemma for \texttt{dyn}].\\
\begin{alltt}
Theorem dyn_equation :
  {forall x} \framebox{i}, {bisimilar_s (dyn x} \framebox{i}{)}
    ({match P x} \framebox{as b return P x = b -> infinite_dyn x -> str} {with}
    {| true =>} \framebox{fun t i} => {SCons (h x) (dyn (g x)}\framebox{(dyn_step1 x t i)}{)}
    {| false =>} \framebox{fun t i =>} {dyn (g' x)} \framebox{(dyn_step2 x t i)}
    {end} \framebox{(refl_equal (P x)) i}).
\end{alltt}
\end{theorem}
We prove this lemma in \cite{BertotKomendaCoq08}, and
give similar proofs for {\tt e\_filter} and {\tt filter}.

In \cite{Bert05}, recursive equations were not given or proved: the
properties of interest for the filter output were directly stated and
proved, culminating with the proof that the infinite stream of prime
integers could be obtained by repetitively filtering from the infinite
stream of natural numbers.  In this paper, we take a more systematic
approach, with the hope that all steps can be automated.

\section{Conclusions and Further Work}\label{sec:concl}
In this paper, we revisited the technique developed in \cite{Bert05}
to model filter functions on streams in a type-theory with
coinductive types and guarded corecursion.  We showed that the same technique
could be applied to a wide class of functions: firstly, we showed that it
could be applied even if the output data-type was not a stream type, and
secondly, we showed that the input data does not have to be
of coinductive type.  In the process, we delineated the various steps of the
description and we showed that the technique can lead to the main
theorem stating that the model exhibits the expected behaviour. 

 In practice,
this work extends the expressive power of guarded co-recursion by
showing that some class of totally and partially productive corecursive functions can be
modelled, even though their initial specification would be expressed
in a non-guarded corecursive equation. The method presented in this paper does not 
cover the productive functions that fail to satisfy the guardedness condition ** of
Section \ref{sec:strguard}. Such functions were studied (e.g.) in \cite{Abel06phd,EGHIK07}. 
Combining these methods with the method we have described here can be one of the objectives for the future work.   

Many of the steps in the technique we describe here can easily be
automated, others are very tricky to formulate.  In future work, we wish
to give a precise description of the class of functions for which the
technique works, and a precise description of each step that is
taken in producing the {\tt eventually} predicate, the first recursive
component using recursion on the ad-hoc predicate {\tt eventually}, etc.
In the end, all the steps could be implemented as a program that takes
as input the syntax tree describing the function one wants to model and
produces both the Coq model and the main lemmas about this model.

Another extension of this work is to study what can be done for functions
that are {\em provably} totally productive (i.e., productive on every
input) but yet not guarded.  In this case, the final recursive equation
should be expressible without proof components.  This extension can
probably take inspiration from the work done on well-founded induction.

\bibliographystyle{abbrv}
\bibliography{Coq}

\end{document}